\documentclass[pre,twocolumn,groupedaddress,showpacs,floatfix]{revtex4}
\usepackage{graphicx}
\usepackage{dcolumn}
\usepackage{amsmath}
\usepackage{bm}
\begin{document}
\title{Drag on particles in a nematic suspension by a moving nematic-isotropic
interface}
\author{John L. West}
\author{Anatoliy Glushchenko}
\author{Guangxun Liao}
\affiliation{Liquid Crystal Institute, Kent State University,
Kent, Ohio 44242}
\author{Yuriy Reznikov}
\affiliation{Institute of Physics of National Academy of Science,
Prospect Nauki 46, Kyiv 252022, Ukraine}
\author{Denis Andrienko}
\affiliation{Max Planck Institute for Polymer Research,
Ackermannweg 10, D-55128 Mainz, Germany}
\author{Michael P. Allen}
\affiliation{Department of Physics and Centre for Scientific Computing,
University of Warwick, CV4 7AL, United Kingdom}
\date{\today}
\begin{abstract}
  We report the first clear demonstration of drag on
  colloidal particles by a moving nematic-isotropic interface.
  The balance of forces explains our observation of periodic,
  strip-like structures that are produced by the movement of these
  particles.
\end{abstract}
\maketitle

Colloidal dispersions of small particles in nematic liquid
crystals are a novel, interesting type of soft matter. The
difference from ordinary colloids arises from the orientational
ordering of the liquid crystal molecules and the resulting
structure in the colloid. Topological defects
\cite{lubensky.tc:1998.a,stark.h:1999, 2001.b,2002.b} and additional long-range
forces between the colloidal particles \cite{lev.bi:1999.a} are
immediate consequences of this ordering. The nematic-induced
interparticle interaction brings a new range of effects to the
system: supermolecular structures
\cite{poulin.p:1997.a,poulin.p:1998.a,loudet.jc:2000.a,loudet.jc:2001.a}, 
cellular structures \cite{anderson.vj:2001.a, anderson.vj:2001.b}, 
and even a soft solid \cite{meeker.sp:2000.a} can be observed. Colloidal
dispersions in liquid crystals also have a wide variety of
potential applications \cite{russel.wb:1989.a}.

A range of problems similar to those of polymer dispersed liquid
crystals also arise in nematic colloidal dispersions. The nematic
ordering makes it difficult to suspend small particles in a liquid
crystal host. Particles often segregate into agglomerates
distributed nonuniformly in the cell. The resulting spatial
distribution of the particles is difficult to control. Our
research explores the factors that affect the spatial
distribution of these particles and indicates ways to control the 
complex morphology of these systems.

In this paper we report the first demonstration of drag on
colloidal particles by a moving nematic-isotropic (NI) interface.
We calculate a critical radius above which the particles cannot be
captured by the moving interface. We predict that this critical
radius is sensitive to the viscous properties of the host liquid
crystal, the value of the anchoring coefficient of the liquid
crystal on the particle surface, and the velocity of
the moving interface. Most important, we can move particles of
specified radius and can control the spatial distribution of these
particles in the cell.

In order to understand how the particles are moved by the
nematic-isotropic transition front we used particles of different
size as well as particles made of different materials.  In the
first part of our experiments we used nearly monodisperse spheres
of silica ($R = 0.005, 0.5$, and $1 \mu \text{m}$). To prove our
predictions based on initial experimental observations and to
demonstrate the controllability of the spatial distribution of
particles in anisitropic colloidal suspensions, we used large
polymer particles, $R={8 \mu \text{m}}$
\footnote{these particles are widely used as spacers in LCD industry
Micropearl SP, made of cross-linked copolymer with divinylbenzene
as a major component, $\rho = 1.05-1.15 \text{g cm}^{-3}$.}. In
all cases particles were dispersed at concentrations of $\phi =
1-5 \text{wt \%}$, in the liquid crystal 5CB at room temperature
($25^0 {\rm C}$) (the isotropic-nematic transition temperature of
pure 5CB is $T_{\rm NI} \approx 35^0 {\rm C}$). The sample was
subjected to ultrasound in order to uniformly disperse the
particles in 5CB. Some of the preparations were made at
higher temperatures, in the nematic or isotropic state of the
liquid crystal.

\begin{figure}
\includegraphics[width=8.5cm]{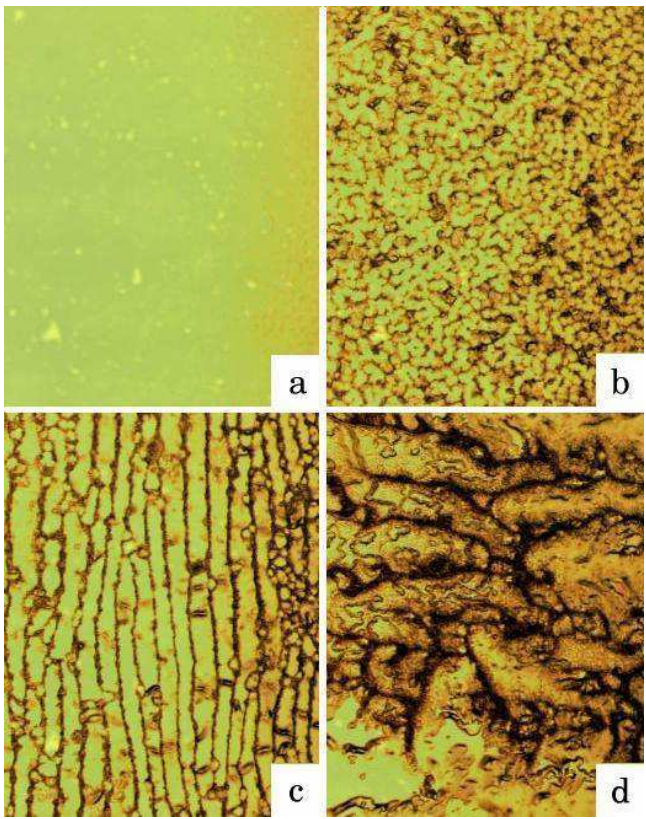}
\caption[Polarized microscopy images of the structures.]{
Polarized microscopy images of different structures, depending on
the cooling rate:
(a) colloidal particles are dispersed homogeneously in the isotropic phase;
(b) {\em cellular} structure:  cooling rate $10^0 \text{C min}^{-1}$;
(c) {\em stripes}: cooling rate $0.1^0 \text{C min}^{-1}$, velocity
of the interface $v \approx 3 \mu \text{m sec}^{-1}$;
(d) {\em root} structure: cooling rate $0.01^0 \text{C min}^{-1}$,
 $v \approx 0.5 \mu \text{m sec}^{-1}$.
Silica particles, $R =  0.005 \mu \text{m}$,
concentration $\phi = 1 \text{wt \%}$.
The NI interface is moving from the left to the right. 
The long side of the images is $1 \text{mm}$.
}
\label{fig:1}
\end{figure}

The homogeneous mixture was deposited between two
polymer-covered glass substrates and heated above the
nematic-isotropic transition point. The cell thickness was 
$100 \mu \text{m}$. The homogeneous suspension was
observed under crossed polarizers using an optical microscope,
Fig.~\ref{fig:1}a.

The mixture was cooled to a temperature below the transition
point. Depending on the rate of cooling, we observed different
structures. Fast quenching to room temperature (cooling rate $10^0
\text{C min}^{-1}$) resulted in phase separation and formation 
of a cellular structure, with
particle-free nematic domains separated by particle-rich regions,
Fig.~\ref{fig:1}b. Properties of these structures have been
reported previously
\cite{anderson.vj:2001.a,anderson.vj:2001.b,meeker.sp:2000.a}.
Decreasing the cooling rate, we observed formation of a striped
structure (Fig.~\ref{fig:1}c). The particle-rich regions were no
longer forming a cellular structure but were arranged in a set of
stripes, separated by particle-free regions. 
Using optical microscope images we postulate that we have
large nematic and isotropic domains separated by a moving
interface. The direction of the stripes is parallel to the moving
interface (the interface was moving from the left to the right of
the cell in the geometry depicted in Fig.~\ref{fig:1}). The
spatial period of the striped structure depended on the cooling
rate, as well as on the particle size. Increasing the particle
size, as well as decreasing the cooling rate, resulted in an
increase of the spatial period. We also noticed that the stripes
do not appear if we have considerably larger silica particles, $R
> 0.5 \mu \text{m}$. Also, decreasing the cooling rate resulted
in a chaotic merging of stripes and formation of a ``{\em
root}''-like pattern, Fig.~\ref{fig:1}d. These results indicate
that the particles are pushed by the moving nematic-isotropic
phase transition front.

\begin{figure}
\includegraphics[width=8.5cm]{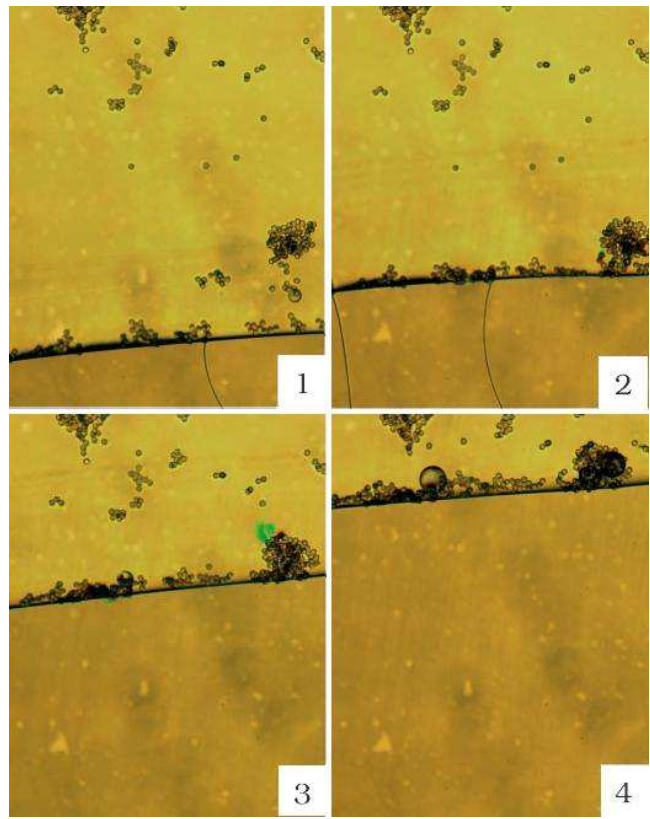}
\caption[Snapshots of the moving nematic-isotropic interface.]{
Snapshots of the moving nematic-isotropic interface made between
parallel polarizers taken with the time interval $\Delta t = 6 \text{min}$.
The nematic phase is on the bottom part of the picture.
Polymer particles, $R={8 \mu \text{m}}$. One can observe
thickening of the particle-rich phase because of the capturing of
the particles by the interface. Velocity of the interface is $v
\approx 3 \mu\text{m sec}^{-1}$. The long side of the images is $1
\text{mm}$.
}
\label{fig:2}
\end{figure}

Using optical microscopy we were able to directly observe the
movement of larger ($R={8 \mu \text{m}}$) but less dense polymer
spheres. We could see these particles being moved by an advancing
nematic to isotropic phase boundary. Fig.~\ref{fig:2} shows
pictures of this moving front taken at different times. Clearly
the particles are pushed by this advancing front, remaining in the
isotropic phase.

A simple analysis  accounts for these experimental observations.
We consider several mechanisms which contribute to the total drag
force acting on a particle at a NI interface. First, the surface
tension coefficient might differ at the particle-nematic or
particle-isotropic part of the interface. An additional pressure
caused by the curvature can be given by $P=2\sigma /R$, where
$R$ is the radius of the curvature (in our case it is the radius
of the particle) and $\sigma $ is the surface tension coefficient.
This pressure contributes to the total drag force as $F_{\sigma
}=2\pi ( \sigma _{N}-\sigma _{I}) R [ 1-( d/R) ^{2}] $, with the
amplitude growing linearly with the droplet radius. Here $d$ is
the distance from the particle center to the interface. The value
of $\Delta \sigma =  \sigma_{N}-\sigma _{I}$ depends on the
surface treatment of the particles and is unknown (and difficult
to measure) for our system. However, the order of magnitude can be
estimated from the change of the surface tension coefficient of
the glass-5CB interface \cite{tintaru.m:2001.a}, $\Delta \sigma
\approx 10^{-2}-10^{-3} \text{dyne cm}^{-2}$.

Second, the particle creates long-range distortions of the
director in a nematic phase. To minimize elastic distortion
energy, the nematic tries to expel particles into the isotropic
phase. The elastic forces have two origins: due to the director
deformations in the bulk nematic and due to the anchoring of the
director at the particle surface. An estimation of these
contributions can be done by dimensional analysis. For the surface
contribution, the only combination which has dimension of force is
$WR$, where $W$ is the anchoring coefficient. Therefore, the
surface contribution to the drag force is proportional to $WR$,
$F_{\text{s}}=WRg_{\text{s}}( d/R)$, where $g_{\text{s}}(x) $ is a
dimensionless function of the penetration depth, $d/R$.

One can have two different situations for the bulk contribution.
For {\em weak} anchoring, $WR/K << 1$, the bulk contribution is
proportional to the squared characteristic deviation of
the director, $\beta_{0}\sim WR/K$ \cite{kuksenok.ov:1996.a}. Now
$W^{2}R^{2}/K$ has the dimension of force, yielding $F_{\text{b}}=
W^{2}R^{2}/K g_{\text{b}}(d/R)$. In contrast, in the case of
strong anchoring, $WR/K >> 1$, the anchoring does not enter the
elastic contribution, and $F_{\text{b}}= K g_{\text{b}}(d/R)$.
Here, again, $g_{\text{b}}(x)$ is a dimensionless function. For
5CB
\footnote{typical values for 5CB (room temperature):
$K_{11} = 6.4 \cdot 10^{-7} \text{dyne}$,
$K_{22} = 3 \cdot 10^{-7} \text{dyne}$,
$K_{33} = 10 \cdot 10^{-7} \text{dyne}$,
$W \approx 10^{-2} - 10^{-4} \text{dyne}$,
$\sigma \approx 40 \text{dyne cm}^{-2}$,
$\Delta \sigma \approx 10^{-2}-10^{-3} \text{dyne cm}^{-2}$ \cite{tintaru.m:2001.a},
$\eta = 0.81 \text{P}$,
density of the silica particles, $\rho = 2.5 \text{g cm}^{-3}$,
density of the polymer, $\rho = 1.05-1.15 \text{g cm}^{-3}$,
packing fraction of the aggregates, $\phi = 0.3-0.8$.},
and typical values of the anchoring energy, $W \approx
10^{-3}-10^{-4} \text{dyn cm}^{-1}$, $WR/K << 1$ for silica
particles and  $WR/K \approx 1$ for polymer particles. Therefore,
for silica particles, we have the weak anchoring regime. In
contrast, polymer particles provide strong anchoring of the
director. We also note that, when particles agglomerate, the
effective radius increases and we have a strong anchoring regime
even for small particles.

Finally, there is a friction drag contribution, which, in
the first approximation, is given by the Stokes formula,
$F_{\eta} = -6\pi R \eta v$ \cite{billeter.jl:2000.a}.
The total drag on the particle is the sum of all contributions,
$F_{\text{drag}} =  F_{\sigma } + F_{\text{b}} + F_{\text{s}} + F_{\eta}$.

Solution of Newton's equations of motion with $F_{\text{drag}}$ as a force
completes the description
of the particle dynamics. It is clear, however, that small heavy
particles cannot be moved by the interface.
The maximal radius can be estimated from the conservation of linear
momentum. To capture a particle of mass $m$, the interface has
to transfer to it a linear momentum $mv$.
If we assume that the particle does not
move (or it moves much slower than the interface, which is valid for
massive particles) then the total linear momentum transferred to the
particle reads
\begin{equation}
m v =
\int_{t_{1}}^{t_{2}}F_{\text{drag}}dt =
\frac{1}{v}\int_{-R}^{R}F_{\text{drag}}\left(
x\right) dx,
\end{equation}
Here we assumed that the interface
touches the particle at time $t_{1}$ and leaves it at time $t_{2}$,
$x=vt$. Substituting $F_{\text{drag}}$ we obtain
\begin{equation}
R_{\max }=\frac{\frac{8}{3}\pi \Delta \sigma
+ \delta _{\text{s}}W
- 6\pi \eta \Delta r}
{\frac{4}{3}\pi \rho v^{2}-\delta _{\text{b}}W^{2}/K},
\label{eq_Rmax}
\end{equation}
where $\delta_i$ are geometrical constants,
$\rho$ is the density of the particle, $\Delta r$ is the final
displacement of the particle due to the drag force.

Several important conclusions can be drawn. First, if the particle
is too big, the moving interface is not able to transfer
sufficient linear momentum to it. Only particles with $R<R_{\max
}\left( v,W,\sigma \right) $ will be captured by the interface.
From eqn~(\ref{eq_Rmax}) one can see that $R_{\max }\propto
v^{-2}$, i.e. only a slowly moving interface is able to capture
the particles. The estimate of this velocity is given by the zero
of the denominator of eqn~(\ref{eq_Rmax}), $v \approx
W/\sqrt{K\rho} \sim 1 \text{mm sec}^{-1}$.
This is of the order of the limiting velocity for the
cellular structure we observed in our experiments: if the
interface moves more slowly, then stripes appear, otherwise the
cellular structure forms (see Fig.~\ref{fig:1}).

The main conclusion is that
$R_{\max }$ is a function of the material parameters, i.e. can be effectively
controlled, for example, by changing the surface treatment of the particles
(anchoring energy $W$). Increase in the anchoring energy leads
to an increase of $R_{\max}$. Moreover, strong enough anchoring favors
formation of a defect near the particle
\cite{poulin.p:1997.a,lubensky.tc:1998.a},
contributing to an even higher energetic barrier created by elastic forces.

On the other hand, if the particle is captured by the interface,
the elastic force scales as $R^{2}$, and the opposing viscous drag scales as
$R$. Therefore, there is a minimal radius, $R_{\min }$,
starting from which particles will be dragged by the interface.
If the particle is dragged by the interface at a constant speed, then
$F_{\text{drag}} = 0$, yielding
\begin{equation}
R_{\min }=\frac{6\pi \eta v-2\pi \Delta \sigma
-\gamma_{\text{s}}W}{\gamma_{\text{b}}W^{2}/K},
\label{eq_Rmin}
\end{equation}
where $\gamma_{i}=g_{i}(0)$ are some constants.
Eqn~(\ref{eq_Rmin}) implies that, in order to be moved by
the interface, the particles have to be big enough. Only in this case
can elastic forces overcome viscous drag. Substituting values typical for
5CB and using the slowest cooling rate, we obtain
$R_{\min} \approx 0.01\mu \text{m}$ which
qualitatively agrees with the minimum size of silica particles
we were able to move.

To explain formation of the striped structure, we note that,
in practice,  particles aggregate into clusters. While an
aggregate moves, it captures more and more particles, growing in size.
The anchoring parameter $WR/K$ also increases and we switch from the
weak anchoring to the strong anchoring regime.
The bulk elastic contribution is then proportional to the
elastic constant $K$ and the elastic force is no longer growing as
$R^2$. Therefore, at some $R_{\text{c}}$, the friction drag overcomes
the elastic contribution and the aggregate
breaks through the interface. A stripe forms and the
particles start to accumulate again.
The condition $F_{\text{drag}} = 0$ gives the critical size of the aggregate
\begin{equation}
R_{\text{c}} = \frac{\gamma_{\text{b}}K}{6\pi\eta v - 2\pi\Delta \sigma},
\end{equation}
which is about $1 \mu \text{m}$ for typical experimental values.

From conservation of mass one can show that the radius of the
aggregate increases linearly with time until it reaches $R_{\text{c}}$,
\begin{equation}
 R=R_{0}+ \frac{1}{4}\phi \frac{\rho_{\text{LC}}}{\rho _{\text{a}}} vt.
\end{equation}
Here $R_{0}$ is the initial radius of the aggregate, $\rho _{\text{a}}$ is the
density of the aggregate. Therefore, the distance
between two stripes is given by
\begin{equation}
\lambda  \approx 4 \phi^{-1}
\frac{\rho _{\text{a}} }{\rho_{\text LC}}  R_{\text{c}},
\end{equation}
and is of the order of $0.1 \text{mm}$, again in qualitative agreement
with experiment, Fig.~\ref{fig:1}.

In conclusion, we have demonstrated moving colloidal particles by
a moving nematic-isotropic interface. We have also determined the
factors such as particle size, anchoring energy and speed of the
moving front, that control the particle movement. By controlling
these factors we can control the morphology of the colloid and its
physical properties. Such control is necessary to develop and
optimize these colloids for specific applications.

\begin{acknowledgments}
This research was supported through INTAS grant 99-00312 and ALCOM
grant DMR 89-20147.
\end{acknowledgments}


\end{document}